\documentclass[apj]{emulateapj}

\input{epsf}

\usepackage{amsmath}

\def\***#1{\textsf{#1}}

\def\M324{\ensuremath{M_{g,324}}}

\newcommand{\notea}{\rlap{\textsuperscript{a}}}
\newcommand{\noteb}{\rlap{\textsuperscript{b}}}
\newcommand{\notec}{\rlap{\textsuperscript{c}}}
\newcommand{\noted}{\rlap{\textsuperscript{d}}}
\newcommand{\notee}{\rlap{\textsuperscript{e}}}
\newcommand{\notef}{\rlap{\textsuperscript{f}}}
\def\p0{\phantom{0}}

\begin{document}

\shorttitle{EVOLUTION OF CLUSTER SCALING RELATIONS}
\shortauthors{KOTOV \& VIKHLININ}
\slugcomment{The Astrophysical Journal, 2005 November}

\title {\emph{XMM-Newton} OBSERVATIONS OF EVOLUTION OF CLUSTER X-RAY
  SCALING RELATIONS AT $z=0.4-0.7$}

\author{O. Kotov\altaffilmark{1}\altaffilmark{,2} and A. Vikhlinin\altaffilmark{2}\altaffilmark{,1}}

\altaffiltext{1}{Space Research Institute, Moscow, Russia}
\altaffiltext{2}{Harvard-Smithsonian Center for Astrophysics, 60 Garden
  St., Cambridge, MA 02138}

\begin{abstract}
  We present a spatially-resolved analysis of the temperature and gas
  density profiles of galaxy clusters at $z=0.4-0.7$ observed with
  \emph{XMM-Newton}. These data are used to derive the total cluster
  mass within the radius $r_{500}$ without assuming isothermality, and
  also to measure the average temperature and total X-ray luminosity
  excluding the cooling cores. We derive the high-redshift $M-T$ and
  $L-T$ relations and compare them with the local measurements.  The
  high-redshift $L-T$ relation has low scatter and evolves as $L\propto
  (1+z)^{1.8\pm0.3}$ for a fixed $T$, in agreement with several
  previous \emph{Chandra} and \emph{XMM-Newton} studies (Vikhlinin et
  al., Lumb et al., and Maughan et al.).  The observed evolution of the
  $M-T$ relation follows $M_{500} \propto E(z)^{-\alpha}$, where we
  measure $\alpha=0.88\pm0.23$. This is in  agreement with
  predictions of the self-similar theory, $\alpha=1$.
\end{abstract}


\section{Introduction}

Scaling relations between the global cluster parameters such as total
mass, X-ray luminosity, and average temperature are important tools for
studies of galaxy clusters and their cosmological applications. Simple
self-similar theory predicts that these relations have a power-law form,
$M\propto T^{3/2}$, and $L_{\text{bol}} \propto T^{2}$ (e.g.,
Kaiser~1991). Deviations of the observed relations from these
theoretical expectations have been used to assess the role of
non-gravitational processes in the cluster formation (see Voit 2005 for
a recent review).

Of prime interest are relations between the total cluster mass and
easily observed quantities, such as the average temperature of the
intracluster medium (ICM) or X-ray luminosity. Such relations allow one
to estimate mass functions for large samples of poorly observed clusters
and use them for cosmological constraints. To derive the mass-observable
relation requires accurate mass measurements in a representative sample
of clusters. Establishing an accurate normalization of the $M-T$ and
$M-L$ relations has been a focus of many recent observational and
theoretical studies
\citep{2000ApJ...532..694N,1996ApJ...469..494E,2001A&A...368..749F,2002ApJ...567..716R,2001ApJ...546..100M,2003MNRAS.345.1241S,2004MNRAS.348.1078B,Vikhlinin05a,Kravtsov05}.

X-ray observations of dynamically relaxed clusters can be used to infer
the total mass via the hydrostatic equilibrium equation (e.g. Sarazin
1988). However, these are technically challenging observations that
require accurate determinations of the cluster temperature profiles.
\mbox{X-ray} mass measurements at radius $r$ are only as accurate as $T$
and $dT/dr$ at that $r$. Simplifying assumptions, such as that the
temperature is constant and gas density follows a $\beta$-model, lead to
large biases in the mass determination \citep{1997ApJ...491..467M}.
Therefore, it is essential to have direct
measurements of the ICM density and temperature distributions at large
radii.  Long-exposure \emph{Chandra} and \emph{XMM-Newton} observations
now provide such measurements for samples of low-redshift clusters
\citep{2005ApJ...628..655V,astro-ph/0502210}, and the resulting $M-T$
relations are in very good agreement with the state-of-the-art
cosmological simulations \citep{2004MNRAS.348.1078B,Kravtsov05}.

In addition to normalization of the local scaling relations,
cosmological applications require experimental constraints on their
evolution at high redshifts. Several studies over the past few years
have used \emph{Chandra} and \emph{XMM-Newton} observations of small
samples of distant clusters to study evolution of the $M-T$ relation
\citep{2004A&A...417...13E,maughan05}.  However, these works have
used an isothermal $\beta$-model to infer cluster masses, which can bias
results on evolution of the scaling relations.

In this \emph{Paper}, we present a spatially-resolved analysis of
\emph{XMM-Newton} observations of a sample of 10 distant clusters
spanning a range of temperatures and redshifts, $2.5<T<9$~keV,
$0.4<z<0.7$. The large effective area of \emph{XMM-Newton} provides good
statistical quality for most of the distant cluster data. The finite
angular resolution of the \emph{XMM-Newton} mirrors ($\sim 4 ''$ FWHM or
24~kpc at $z=0.5$) is the main technical challenge in this analysis.
However, this problem can be solved and deconvolved ICM temperature and
density profiles can be restored for our distant clusters. We use these
measurements to infer the total cluster mass, as well as the X-ray
luminosity and temperature excluding the central cool regions, and thus
study the evolution of cluster scaling relations at $z>0.4$.

All distance-dependent quantities are derived assuming the
$\Omega_M=0.3$, $\Omega_\Lambda=0.7$ cosmology with the Hubble constant
$H_0=71$km~s$^{-1}$~Mpc$^{-1}$. Statistical uncertainties are quoted at
$68\%$ CL.

\section{Observations and Data Reduction}

\begin{deluxetable*}{p{4.0cm}clclcl}
\tablecolumns{7} 
\tablewidth{0pc} 
\tablecaption{Summary of \emph{XMM-Newton} observations\label{live}}
\tablehead{ 
\colhead{Name}    & \colhead{$t_{\text{PN}}$\notea} &   \colhead{$\delta_{\text{PN}}$\noteb}   & \colhead{$t_{\text{MOS1}}$\notea} &
\colhead{$\delta_{\text{MOS1}}$\noteb} & \colhead{$t_{\text{MOS2}}$\notea} & \colhead{$\delta_{\text{MOS2}}$\noteb} }
\startdata 
CL 0016+16\dotfill        & 20.5  & 0.86      &    25.3          &     0.94         &     22.2        &    0.90 \\
CL 0024+17\dotfill        & 31.7  & 0.92      &    43.6          &     0.89         &     42.4        &    0.88 \\
MS0302.5+1717\dotfill     & \p05.7& 1.20      &    \p09.9        &     1.00         &     \p09.9      &    0.96 \\
MS1054.4-0321\dotfill     & 12.5  & 1.07      &    19.8          &     0.97         &     15.8        &    0.97 \\
RXJ1120.1+4318\dotfill    & 11.6  & 0.96      &    16.1          &     0.99         &     14.6        &    0.96 \\
RXJ1334.3+5030\dotfill    & 20.0  & 1.01      &    26.5          &     0.99         &     26.5        &    0.97 \\
WJ1342.8+4028\dotfill     & 17.8  & 1.07      &    27.6          &     1.02         &     25.3        &    1.01 \\
WARPJ0152.7-1357\dotfill  & 33.4  & 0.98      &    44.7          &     0.94         &     46.8        &    0.90 \\
RXJ0505.3+2849\dotfill    & 12.2  & 0.99      &    21.5          &     0.93         &     20.8        &    0.90 \\
CL0939+472\dotfill        & 20.7  & 0.91      &    30.5          &     0.85         &     30.5        &    0.82 
\enddata
\tablenotetext{a}{~Clean exposure times, ksec.} \tablenotetext{b}{~Ratio of observed 10--15~keV flux outside the field of view to that
  in the ``closed dataset''.}
\end{deluxetable*}

Our sample was selected from publicly available \emph{XMM-Newton}
observations of distant clusters.  The final goal was to measure the
cluster scaling relations at high redshift and we restricted our sample
to objects with $z \gtrsim 0.4$. The selected clusters are listed in
Table~\ref{live}.

We used the data from all EPIC cameras (MOS1, MOS2, and PN). 
For data reduction, we used the {\em XMM-Newton} Science Analysis System
(SAS) v6.0.0 and the calibration database with all updates available
prior to November, 2004. All data were reprocessed with these latest
versions of gain files. The initial data screening was applied using
the recommended sets of event patterns, 0--12 and 0--4 for the MOS and PN
cameras, respectively, and  excluding all known bad CCD pixels. 
Our science goals required accurate background subtraction. The EPIC
background is highly variable and only its quiescent component can be
accurately modeled. Therefore, our next step was to detect and
exclude the periods of flaring background. We extracted the light curves
for each camera in the 2--15~keV band using the data from the entire
field of view excluding detectable sources. The light curves were binned
to 200~s time resolution and flares were detected as $>2\sigma$
deviations from the mean. Experiments with different choices of 
energy bands and flare detection thresholds have shown that our choice
was close to optimal. For example, more flares are typically detected in
the 2--15~keV band than in  the frequently used $>10$~keV band
because some flares are soft. In some observations, almost the entire exposure
was affected by flares, and these observations had to be discarded. 
The clean exposure times for each camera  are listed in Table \ref{live}.

To account for strong \emph{XMM} mirror vignetting, we used an approach
proposed by \cite{Arnaud01}. Each photon was assigned a weight
proportional to inverse vignetting and these weights were then used in
computing images and spectra. This was done using the SAS tool
\emph{evigweight}.

%

Background modeling in our analysis was implemented following the
double-subtraction method of \cite{2002A&A...390...27A}. The first step
of this method is to subtract the particle-induced background
component. This component can be estimated from a set of \emph{ XMM}
observations with the filter wheel closed (so called ``closed data'').
We compiled the closed dataset from public observations available in the
\emph{XMM} data archive; these data were reduced following steps indetical to
those of the science observations. The closed background was adjusted to the
cluster observations using the observed flux in the 10--15~keV band
outside the field of view. The scaling factors are listed in
Table~\ref{live}. The second step is to determine the cosmic X-ray
background (CXB) component. Its spatial distribution should be flat
because vignetting correction is already applied. Therefore, the CXB
spectrum can be measured in the source-free regions of the field of
view, typically at $r=7'-9'$ from the cluster center and then directly
subtracted.

Finally, we computed corrections for photons registered during the CCD
readouts (``out-of-time events''), using the SAS tools \emph{epchain}
and \emph{emchain} with the \emph{withoutoftime=yes} flag.

\section{Image analysis}

\begin{deluxetable*}{cccccccccc}
\tabletypesize{\scriptsize}
\tablecolumns{10}
\tablewidth{0pc}
\tablecaption{Results of Image Analysis\label{main1}}
\tablehead{ 
\colhead{Name}    &  &   \multicolumn{3}{c}{$\beta$ fit} & & \multicolumn{4}{c}{$\alpha-\beta$ fit} \\
\cline{3-5}\cline{7-10}\\
\colhead{} & \colhead{$z$} & \colhead{$\beta$} & \colhead{$r_{c}$, kpc} & \colhead{$\chi^2$/dof} && \colhead{$\alpha$} & \colhead{$\beta$} & \colhead{$r_{c}$, kpc}& \colhead{$\chi^2$/dof} }
\startdata 
CL 0016+16\dotfill            & 0.54 &$0.76\pm^{0.01}_{0.01}$&$267.8\pm^{7.5}_{7.8}$  & 176.4/128 &&$0.64\pm^{0.11}_{0.12}$&$0.84\pm^{0.03}_{0.03}$&$372.7\pm^{32.1}_{27.8}$    & 158.1/127 \\
CL 0024+17\dotfill            & 0.39 &$0.59\pm^{0.02}_{0.01}$&$89.0\pm^{8.0}_{4.8}$   &  98.7/65  &&$1.56\pm^{0.13}_{0.16}$&$0.72\pm^{0.07}_{0.05}$&$254.5\pm^{58.8}_{47.5}$    &  78.3/64  \\
MS0302.5+1717\dotfill         & 0.42 &$0.65\pm^{0.06}_{0.05}$&$118.7\pm^{20.8}_{18.2}$&  74.9/83  &&$1.63\pm^{0.23}_{0.36}$&$1.03\pm^{3}_{0.28}$   &$456.0\pm^{963.7}_{196.6}$  &  69.0/82  \\  
MS1054.4-0321\notea\dotfill   & 0.82 &\nodata                &\nodata                 &  \nodata  && \nodata               & \nodata               &  \nodata                   & \nodata   \\
RXJ1120.1+4318\dotfill        & 0.60 &$0.81\pm^{0.04}_{0.04}$&$203.8\pm^{14.2}_{15.0}$&  77.6/83  &&$0.29\pm^{0.40}_{0.69}$&$0.84\pm^{0.09}_{0.08}$&$231.0\pm^{60.1}_{59.6}$    &  77.4/82  \\
RXJ1334.3+5030\dotfill        & 0.62 &$0.61\pm^{0.02}_{0.02}$&$127.7\pm^{9.4}_{9.4}$  &  94.5/77  &&$1.57\pm^{0.17}_{0.27}$&$0.95\pm^{0.75}_{0.21}$&$520.3\pm^{665.9}_{201.5}$  &  87.2/76  \\
WJ1342.8+4028\dotfill         & 0.70 &$0.49\pm^{0.03}_{0.02}$&$56.3\pm^{15.5}_{10.4}$ &  40.7/56  &&$2.03\pm^{0.15}_{0.21}$&$0.78\pm^{3}_{0.21}$   &$579.1\pm^{1690.5}_{393.46}$&  35.1/55  \\
WARPJ0152.7-1357\notea\dotfill& 0.83 &  \nodata              &      \nodata           &  \nodata  && \nodata               & \nodata               &  \nodata                   & \nodata   \\
RXJ0505.3+2849\dotfill        & 0.51 &$0.69\pm^{0.14}_{0.10}$&$162.8\pm^{46.2}_{38.4}$& 103.1/96  &&$1.07\pm^{0.41}_{0.80}$&$0.92\pm^{**}_{0.26}$  &$353.2\pm^{**}_{167.6}$     & 101.6/95  \\
CL0939+472\notea\dotfill      & 0.41 &  \nodata              &      \nodata           & \nodata   && \nodata               & \nodata               &  \nodata                   & \nodata   
\enddata
\tablenotetext{a}{Profile fitting was not performed for irregular clusters}
\end{deluxetable*}

\begin{deluxetable*}{p{4cm}cccccrccc}
\tabletypesize{\scriptsize}
\tablecolumns{10} 
\tablewidth{0pc} 
\tablecaption{Results of Spectral and Mass Determination\label{main2}}
\tablehead{\colhead{Name}    & \colhead{$T$\notea} &  \colhead{$Z$\noteb} &\colhead{$r_s${}\notec} & \colhead{$T^{spec}_{500}$\noted} &
  \colhead{$T^{emw}_{500}$\notee} &  \colhead{$r_{500}$} &  \colhead{$M_{500}$}  &  \colhead{$L_{\text{bol}}$\notef}\\[3pt] 
 \colhead{} & \colhead{(keV)} &  \colhead{(Z$_{\odot}$)} &\colhead{(Mpc)}  & \colhead{(keV)} & \colhead{(keV)}  & \colhead{(Mpc)} & \colhead{$(10^{14}M_{\sun})$} & \colhead{(erg~s$^{-1}$)}}
\startdata
CL 0016+16\dotfill       &$ 8.9 \pm ^{0.3}_{0.3}$  & $0.17 \pm ^{0.04}_{0.04}$  & 1.0 & $9.3\pm^{0.4}_{0.3}$   & $9.4\pm ^{0.4}_{0.3}$   &  $1.19\pm ^{0.05}_{0.05}$  & $8.83\pm ^{1.08}_{0.98}$ &       $50.79\times10^{44}$\p0 \\
CL 0024+17\dotfill       &$ 3.5 \pm ^{0.1}_{0.1}$  & $0.29 \pm ^{0.06}_{0.06}$  & 0.5 & $3.6\pm^{0.2}_{0.2}$  & $4.0\pm ^{0.3}_{0.3}$    &  $0.74\pm ^{0.02}_{0.02}$  & $1.77\pm ^{0.10}_{0.10}$ &       $3.98\times10^{44}$     \\
MS0302.5+1717\dotfill    &$ 4.5 \pm ^{0.5}_{0.4}$  & $0.61 \pm ^{0.27}_{0.24}$  & 0.5 & $4.1\pm ^{0.8}_{0.8}$  & $4.5\pm ^{0.9}_{0.9}$  &  $0.78\pm ^{0.07}_{0.07}$  & $2.15\pm ^{0.63}_{0.55}$ &       $4.29\times10^{44}$     \\
MS1054.4-0321\dotfill    &$ 7.5 \pm ^{0.7}_{0.5}$  & $0.35 \pm ^{0.09}_{0.09}$  & 0.7 &  \nodata  &  \nodata              &   \nodata               &   \nodata   &       $30.49\times10^{44}$\p0     \\
RXJ1120.1+4318\dotfill   &$ 4.9 \pm ^{0.3}_{0.3}$  & $0.41 \pm ^{0.11}_{0.11}$  & 1.0 & $5.0\pm ^{0.3}_{0.3}$ & $5.1\pm ^{0.4}_{0.3}$  &  $0.94\pm ^{0.07}_{0.07}$  & $4.64\pm ^{1.14}_{0.96}$ &       $13.04\times10^{44}$\p0     \\
RXJ1334.3+5030\dotfill   &$ 4.6 \pm ^{0.4}_{0.3}$  & $0.22 \pm ^{0.14}_{0.13}$  & 0.8 & $4.6\pm ^{0.4}_{0.3}$  & $4.7\pm ^{0.4}_{0.3}$  &  $0.78\pm ^{0.04}_{0.03}$  &  $2.73\pm ^{0.48}_{0.30}$ &       $7.48\times10^{44}$     \\
WJ1342.8+4028\dotfill    &$ 3.5 \pm ^{0.3}_{0.3}$  & $0.56 \pm ^{0.29}_{0.24}$  & 0.6 & $3.1\pm ^{0.3}_{0.4}$  & $3.4\pm ^{0.4}_{0.4}$  &  $0.59\pm ^{0.03}_{0.04}$  & $1.29\pm ^{0.17}_{0.24}$ &       $3.66\times10^{44}$    \\
WARPJ0152.7-1357\dotfill &$ 6.2 \pm ^{0.4}_{0.4}$  & $0.33 \pm ^{0.08}_{0.08}$  & 0.8 &  \nodata               & \nodata                   &   \nodata               &   \nodata                 &       $21.71\times10^{44}$\p0   \\
RXJ0505.3+2849\dotfill   &$ 2.5 \pm ^{0.4}_{0.5}$  & $0.61 \pm ^{0.65}_{0.42}$  & 1.0 &  $2.8\pm ^{0.4}_{0.3}$ & $3.2\pm ^{0.5}_{0.4}$  &  $0.65\pm ^{0.03}_{0.04}$  & $1.37\pm ^{0.23}_{0.21}$ &       $1.58\times10^{44}$  \\
CL0939+472\dotfill       &$ 5.3 \pm ^{0.2}_{0.2}$  & $0.22 \pm ^{0.06}_{0.06}$  & 1.0 & \nodata                & \nodata                   &   \nodata               &   \nodata                 &       $10.05\times10^{44}$       
\enddata
\tablenotetext{a}{~Best-fit temperature to the integral cluster
  spectrum.}  
\tablenotetext{b}{~Best-fit metallicity to the integral cluster
  spectrum.} 
\tablenotetext{c}{~Spectral extraction radius.}
\tablenotetext{d}{Spectroscopic temperature within $70\,\text{kpc}<r<r_{500}$. 
Note that $T^{sp}_{500}$ was renormalized by
  $+8\%$ to account for \emph{Chandra} vs.\ \emph{XMM-Newton}
  cross-calibration.}
\tablenotetext{e}{ Emission-weighted temperature within $70\,\text{kpc}<r<r_{500}$. 
$T^{ew}_{500}$ was renormalized by
  $+8\%$ to account for \emph{Chandra} vs.\ \emph{XMM-Newton}
  cross-calibration.}
\tablenotetext{f}{~Bolometric luminosity within $70\,\text{kpc}<r<1400$~kpc. The
  central region was not excluded for clusters with irregular
  morphology, MS1054.4-0321, WARPJ0152.7-1357,
  CL0939+472. Emission-weighted temperature, $T^{ew}_{500}$, was used to
  compute $L_{\text{bol}}$  for all clusters except  MS1054.4-0321, WARPJ0152.7-1357,
  CL0939+472.}
\end{deluxetable*}

We excluded all detectable point sources from the data in our spectral
and spatial analysis. 
The sources were detected separately in the ``optimal'' 0.3-3~keV,
``soft'' 0.3-0.8~keV, and ``hard'' 2.0-6.0~keV energy bands. Detected
point sources were masked with circles of 80\% PSF power radii.

Spatial  analysis of the cluster emission was performed in the $0.5 - 2.0$ keV energy band.  
The images for each camera were corrected for vignetting and
out-of-time events, with the particle background component subtracted
as described in \S~2, and analyzed independently. We extracted the
azimuthally averaged surface brightness profiles centered on the X-ray
surface brightness peak, even for clusters with irregular morphology
(see below), excluding the CCD gaps and circles around the point
sources. The obtained profiles were used to derive the parameters of
the spatial distribution of the ICM, the cluster fluxes, and the
background levels. 

The cluster surface brightness profiles are often modeled with the so-called 
$\beta$-model, $n_e^2 \propto (1+r^2/r_c^2)^{-3\beta}$ or $S_x
\propto (1+r^2/r_c^2)^{-3\beta+0.5}$ \citep{Cavaliere76}. 
However, this model poorly describes
clusters with sharply peaked surface bigness profiles  related to
the radiative cooling of the ICM in the cluster centers. 
The  simple modification of the $\beta$-model \citep{Pratt02}
\begin{equation}\label{eq:alpha-beta}
n_e^2 \propto \frac{(r/r_c)^{-\alpha}}{(1+r^2/r_c^2)^{3\beta-\alpha/2}},
\end{equation}
allows adequately describing these cooling regions in the 
low redshift clusters. 
For $\alpha=0$, this $\alpha$-$\beta$ model is identical to the
 usual $\beta$-model.

 The model for the observed surface brightness profiles can be obtained
 by numerical integration of eq.(\ref{eq:alpha-beta}) along the line of
 sight and convolution of the result with the XMM PSF\footnote{We used
   the latest available values of the King function parametrization of
   the MOS1,
   MOS2, PN PSF, see\\
   http://xmm.vilspa.esa.es/docs/documents/CAL-TN-0018-2-4.pdf}. To
 represent the uniform sky X-ray background, we added a constant
 component to the model and treated it as a free parameter.  The values
 of $\alpha$, $\beta$, and $r_c$ were derived from the joint fit to the
 observed profiles in the MOS1,MOS2, and PN cameras, with the overall
 normalizations and background levels fitted independently for each camera.
 For comparison, we also fitted the standard $\beta$-model by setting
 $\alpha=0$.  The obtained parameters for the $\alpha$-$\beta$ model and
 the standard $\beta$ model fits are summarized in Table \ref{main1}.

 The best-fit ICM model was used to derive the total flux  and 
the total mass from the hydrostatic equilibrium equation (see \S
 6 below). Strictly speaking, eq.(\ref{eq:alpha-beta}) can not be
 applied to the ICM distribution in clusters with irregular morphology.
 In these cases, the model fit was used only to measure the image
 background, and we do not list the values of $\alpha$, $\beta$, and
 $r_c$ in Table \ref{main1}.

\section{Spectral Analysis}

\begin{figure}
\vspace*{-1.5\baselineskip}
\centerline{
\includegraphics[width=0.99\linewidth]{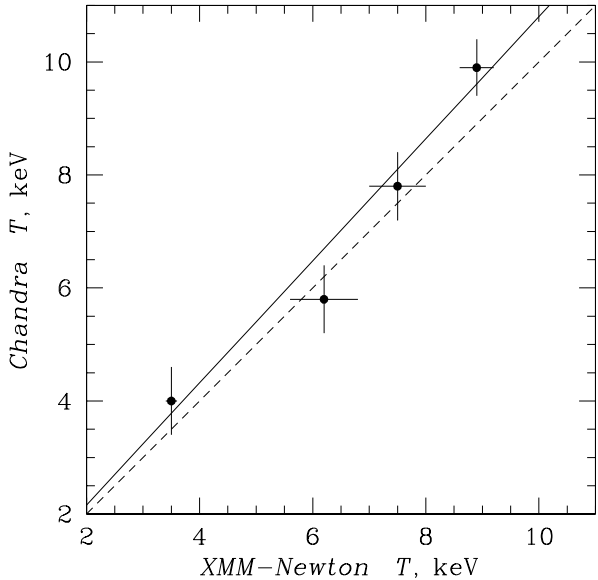}
}
\caption{Comparison of {\em XMM-Newton} and {\em Chandra}
  temperatures. Dashed line corresponds to
  $T_{\text{\emph{XMM}}}=T_{\text{\emph{Chandra}}}$, and solid line
  shows the best-fit relation,
  $T_{\text{\emph{XMM}}}=0.92\,T_{\text{\emph{Chandra}}}$. 
}
\label{fig:chandra_xmm}
\end{figure}

The cluster spectra were extracted in annuli and circles within radii
listed in Table \ref{main2}.  The response matrices and effective area
files were generated by the standard SAS tasks.  Because the data were
previously vignetting corrected, the effective area files were created
for the on-axis position using the routine {\em arfgen}.  The response
matrices were generated in the spectrum extraction region via {\em
  rmfgen}.  The spectra were binned, so that there
were at least 20 counts per bin.  The observed spectra were fitted in the
$0.5 - 10$ keV energy band. We checked that there were no prominent 
residuals near the instrumental lines  and that
exclusion of the instrumental line regions did not affect final results.

We used the Mewe-Kaastra-Liedahl plasma emission model \citep{1985A&AS...62..197M},
with free metallicity and Galactic absorption fixed at the value derived
from the radio surveys \citep{1990ARA&A..28..215D}.  The EPIC-PN, MOS1,
and MOS2 spectra for each extraction region were fitted jointly, with the
spectral parameters tied and relative normalizations free. 
The resulting spectral parameters are listed in Table \ref{main2}. 
Our single-temperature fits to wide-beam spectra are in good agreement
with the published results (e.g., \cite{2004A&A...420..853L}, \cite{2003MNRAS.340.1261W}).

\section{The Cross-Calibration between XMM-Newton and Chandra}

To compare our results with the low-redshift scaling relations we need
to correct for any systematic differences between {\em ASCA}, {\em
  Chandra}, and {\em XMM-Newton} measurements.  \cite{Vikhlinin02}
verified that there is no systematic difference between the {\em
  Chandra} and the {\em ASCA} temperature measurements.  Therefore we
need to cross-calibrate only \emph{XMM-Newton} and \emph{Chandra}.
Several of our clusters were in the \cite{Vikhlinin02} sample.  The
comparison of \emph{XMM-Newton} and \emph{Chandra} temperatures for
these clusters is shown in Fig.~\ref{fig:chandra_xmm}.  There is a good
overall agreement, although the \emph{Chandra} temperature for the
hottest cluster, CL0016+16, is marginally higher than our value. The
linear fit gives $T_{\text{\emph{XMM}}}=(0.92 \pm
0.08)\,T_{\text{\emph{Chandra}}}$. A similar systematic difference is
found in the comparison of the \emph{Chandra} and \emph{XMM} temperature
profiles of nearby clusters \citep{2005ApJ...628..655V}. We applied this
correction factor to all temperature values because we use the
\emph{ASCA} $L-T$ relation as low-redshift reference.  For $M-T$
relation, this correction is unimportant (see \S 7). Comparison of the
cluster luminosities shows a good agreement, within $\pm5$\%, between
\emph{Chandra} and \emph{XMM}.

\section{Temperature profiles and mass modeling}

\begin{figure*}
\centerline{\includegraphics[width=0.43\linewidth]{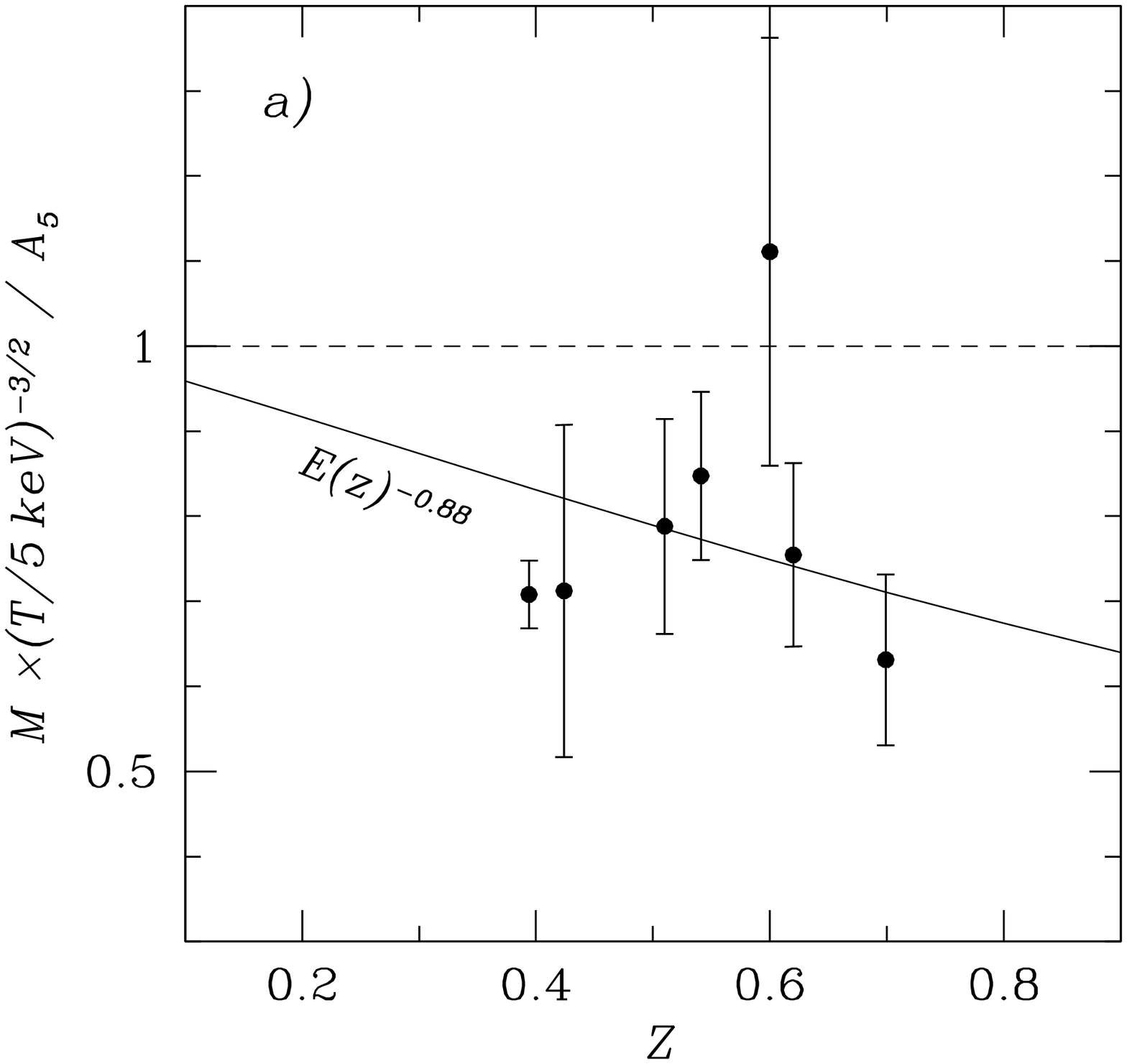}%
  \hfill%
  \includegraphics[width=0.43\linewidth]{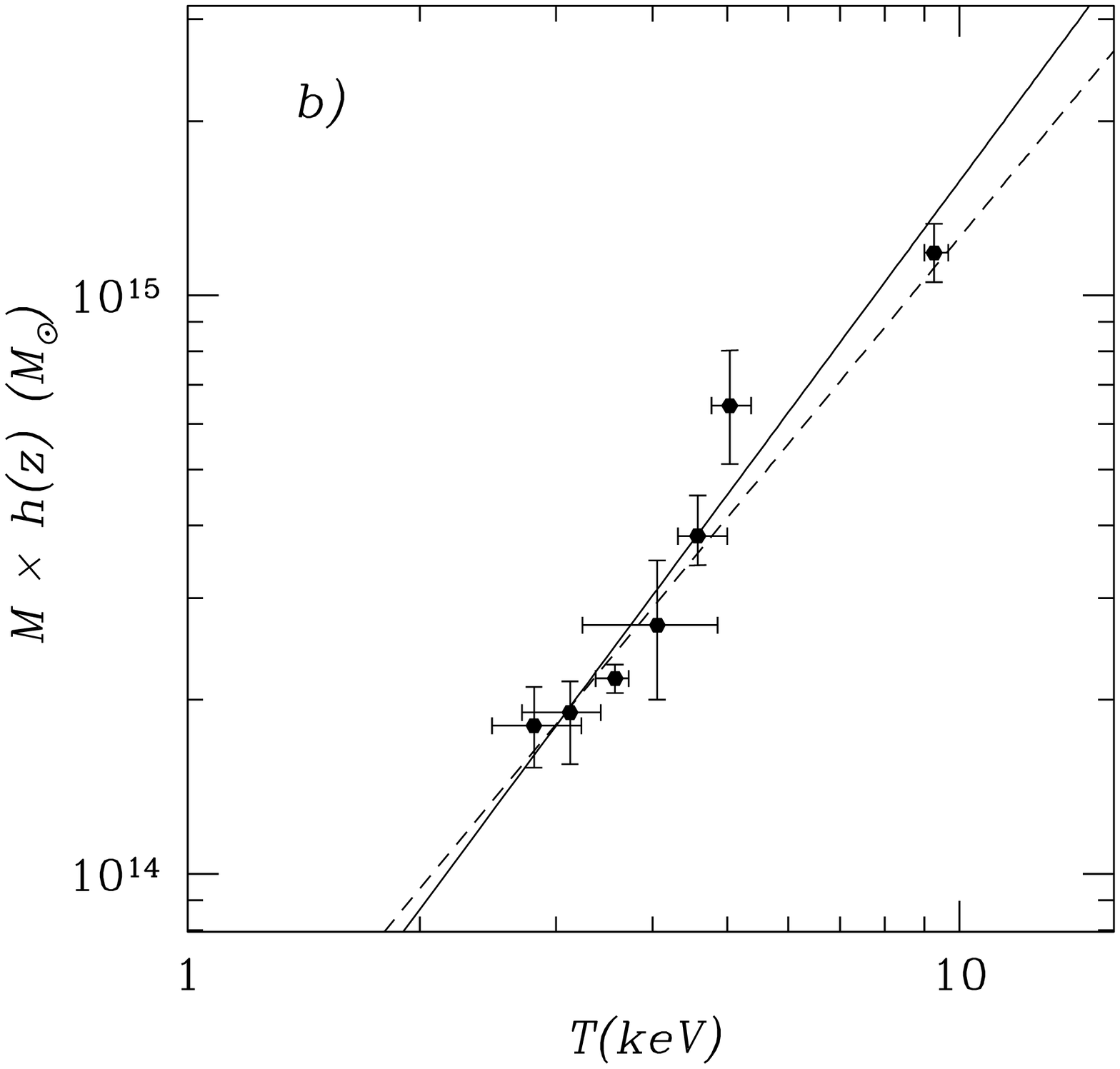}%
}
\caption{Correlation of cluster spectroscopic temperature and total mass. The data are corrected 
for the found {\em XMM-Chandra(ASCA)} systematic discrepancy in the temperature measurement. 
{\em (a)}
The quantity $a = M_{500}/T_{spec}^{1.5}/A_5$ as a function of $z$. The normalization $A_5$ is adopted 
from   \emph{Chandra} observations of 13 low-redshift clusters \citep{Vikhlinin05a}. 
{\em (b)} 
The solid line shows the derived best-fit $M-T$ correlation. 
The dashed line shows the low-redshift result from \cite{Vikhlinin05a}.
}
\label{fig:mt}
\end{figure*}

The statistical quality of \emph{XMM-Newton} data allows to reconstruct
the temperature profiles from all our clusters. The main complication
is that the XMM PSF size is non negligible compared with the angular
size of distant clusters.  For example , $50\%$ of the flux from 
the central 100 kpc region ($15''$ at $z$=0.6) is scattered to
larger radii ($\approx 90\%$ of the flux stays within 260 kpc regions). 
The temperature in these regions is often lower than
the cluster average because of radiative cooling, and hence 
this scattered flux can significantly bias the temperature
measurements at large radii.  

We corrected for the XMM PSF using an approach used for $XMM-Newton$ data analysis by \cite{2004A&A...423...33P}.  
Using the best-fit $\alpha$-$\beta$ models of the cluster brightness and the XMM
PSF calibration, we calculated the redistribution matrix, $R_{ij}$, of
each temperature to each annulus that represents relative contribution
of emission from annulus $i$ to the observed flux in annulus $j$. 
The model spectrum, $S_j$, is then given by 
\begin{equation}
  \label{eq:t:mix}
  S_j=\sum R_{ij} S(T_i),
\end{equation}
where $T_i$ is the temperature in annulus $i$ and $S(T_i)$ is the
\emph{mekal} spectrum for this temperature. Fitting this model to the
observed spectra in all annuli simultaneously and treating all $T_i$ as
free parameters gives the deconvolved temperature profile. The raw and
deconvolved temperature profiles are shown in
Fig.\ref{fig:cl0016}--\ref{fig:ms0302}.\footnote{The prime motivation
  for temperature profile analysis is to derive the total masses from
  the hydrostatic equilibrium equation.  This cannot be done for
  irregular clusters ,\, MS1054.4-0321,\, WARPJ0152.7-1357,\, CL0939+472, so we have not analyzed the
  temperature profiles in these cases.} 
The deconvolved temperature profiles  
are within  1 $\sigma$ of the raw measurements in all cases. 
However the PSF correction is systematic and results in stronger 
temperature gradients. Thus neglecting this effect
can slightly biases the mass measurements.
In many cases, we observe a decrease of temperature at large radii,
which is qualitatively consistent with the results for low-redshift
clusters \citep{1998ApJ...503...77M,2002ApJ...567..163D,2005ApJ...628..655V}. 
The only cluster that appears approximately isothermal is RXJ1120.1; our results 
for this cluster are fully consistent with the analysis by
\cite{2002A&A...390...27A}.  
We fitted  the observed temperature profiles by the function:
\begin{equation}
T(r)=T_0 \frac{1}{(1+(r/r_0)^2)^{\alpha}}
\end{equation}
A similar model describes the temperature profiles for low redshift
clusters \citep{Vikhlinin05a}. 
For local clusters, $r_0$ scales with 
the average temperature as $r_0=0.284\,(T/\text{1\,keV})^{0.537}$~Mpc
(see Fig.~16 in \cite{Vikhlinin05a}). The statistical uncertainties  in our
temperature profiles are insufficient to fit $T_0$, $r_0$, and the outer
slope $\alpha$.  
Therefore we fixed $r_0$ at the value suggested by the low-redshift
correlation, with an additional scaling, $r_0 \propto 1/E(z)$,
to account for the redshift dependence of the virial radius for
a fixed temperature \citep{1998ApJ...495...80B}. 

To fit the observed profiles, we projected the 3-D model along the line
of sight using the emission measure profile from the best fit
$\alpha$-$\beta$ model. Projection was based on a weighting method that
correctly predicts the best-fit spectral $T$ for a mixture of different
temperature components \citep{2004MNRAS.354...10M,Vikhlinin05b}. For
those clusters with the central temperature decrements we excluded the
innermost bin from the fit.  This procedure is correct because our
temperature profiles were corrected for the XMM PSF.  The obtained
best-fit models are shown along with the deconvolved temperature
profiles in Fig.\ref{fig:cl0016}--\ref{fig:ms0302}.

Assuming hydrostatic equilibrium for the ICM, we can use the best fit
temperature and density profiles to derive the total cluster masses:
\begin{equation}
  \label{eq:def1_m}
M(r)=-\frac{rT(r)}{G \mu m_p}\left( \frac{d \log \rho(r)}{d \log r} + \frac{d \log T(r)}{ d \log r} \right)
\end{equation}
The mass was calculated within the radius that corresponds to the 
mean overdensity $\Delta=500$ relative to the critical density
at the cluster redshift.

The uncertainties on the masses were calculated from Monte-Carlo
simulations. The mass uncertainties are dominated by statistical
uncertainties of the temperature profiles, and we neglected all other
sources of error. We used the best-fit temperature profile as a
template, applied Gaussian scatter with the \emph{rms} equal to the
statistical uncertainties, and fitted the simulated profile and
derived the mass. The mass uncertainty was estimated as an \emph{rms}
scatter in 1000 simulations.

For each cluster we calculated the following temperature averages:

$T_{emw}$ -- $T(r)$ weighted with $\rho_{gas}(r)^2$. $T_{emw}$ is
needed to self-consistently compare the high-z $L-T$ relation of
our sample with the low-redshift result of \cite{Markevitch98}.

$T_{spec}$ -- $T_{spec}$ is obtained by integrating a combination of
$T(r)$ and  $\rho_{gas}(r)^2$ as described in  \cite{Vikhlinin05b}.
$T_{spec}$ is needed for comparison with the low-redshift $M-T$
relation from  \cite{Vikhlinin05a}.

$T_{spec}$ and $T_{emw}$  were averaged in the radial
range 70 kpc $<$ r $<$ $r_{500}$.
The uncertainties on $T_{emw}$ and $T_{spec}$ were also calculated from Monte-Carlo 
simulations. The obtained values of $T_{emw}$, $T_{spec}$, $M_{500}$,
and the corresponding overdensity
radius $r_{500}$ are listed in Table \ref{main2}. The $r_{500}$ radii
are also  shown by
vertical dotted lines in Fig.\ref{fig:cl0016}--\ref{fig:ms0302}.

To test how background subtraction can affect our mass result,
we  checked two sources of uncertainties. First, we varied 
normalization of the article-induced component by  $\pm10\%$.
Normalization of CXB component was varied by  $\pm5\%$ (quoted XMM
vignetting uncertainties). We refitted all spectra with renormalized
background and repeated the mass analysis. The resulting variations in
$M_{500}$ were well within our statistical uncertainties.

\section{Evolution of $M-T$ relation}

Self-similar theory (e.g, \cite{1998ApJ...495...80B}) predicts that the
relation between cluster mass and temperature is a power law that
evolves as $M_{\Delta}/T^{3/2} \propto E(z)^{-1}$, where
$E(z)=H(z)/H_0=(0.3(1+z)^3+0.7)^{1/2}$ for the adopted cosmology.  Our
measurements for distant clusters can be used to test these predictions. 

\begin{figure*}
\centerline{\includegraphics[width=0.43\linewidth]{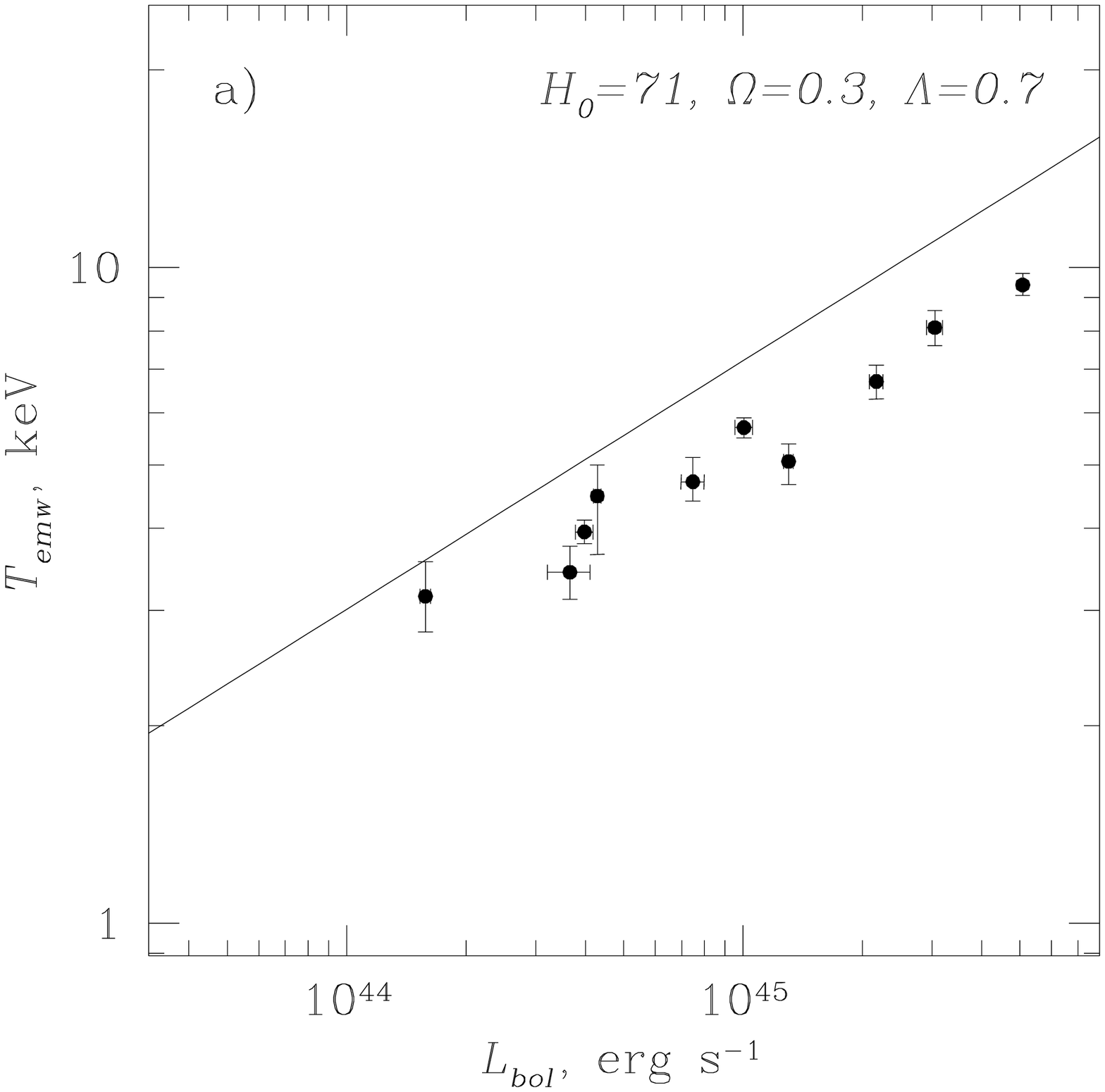}%
  \hfill%
  \includegraphics[width=0.43\linewidth]{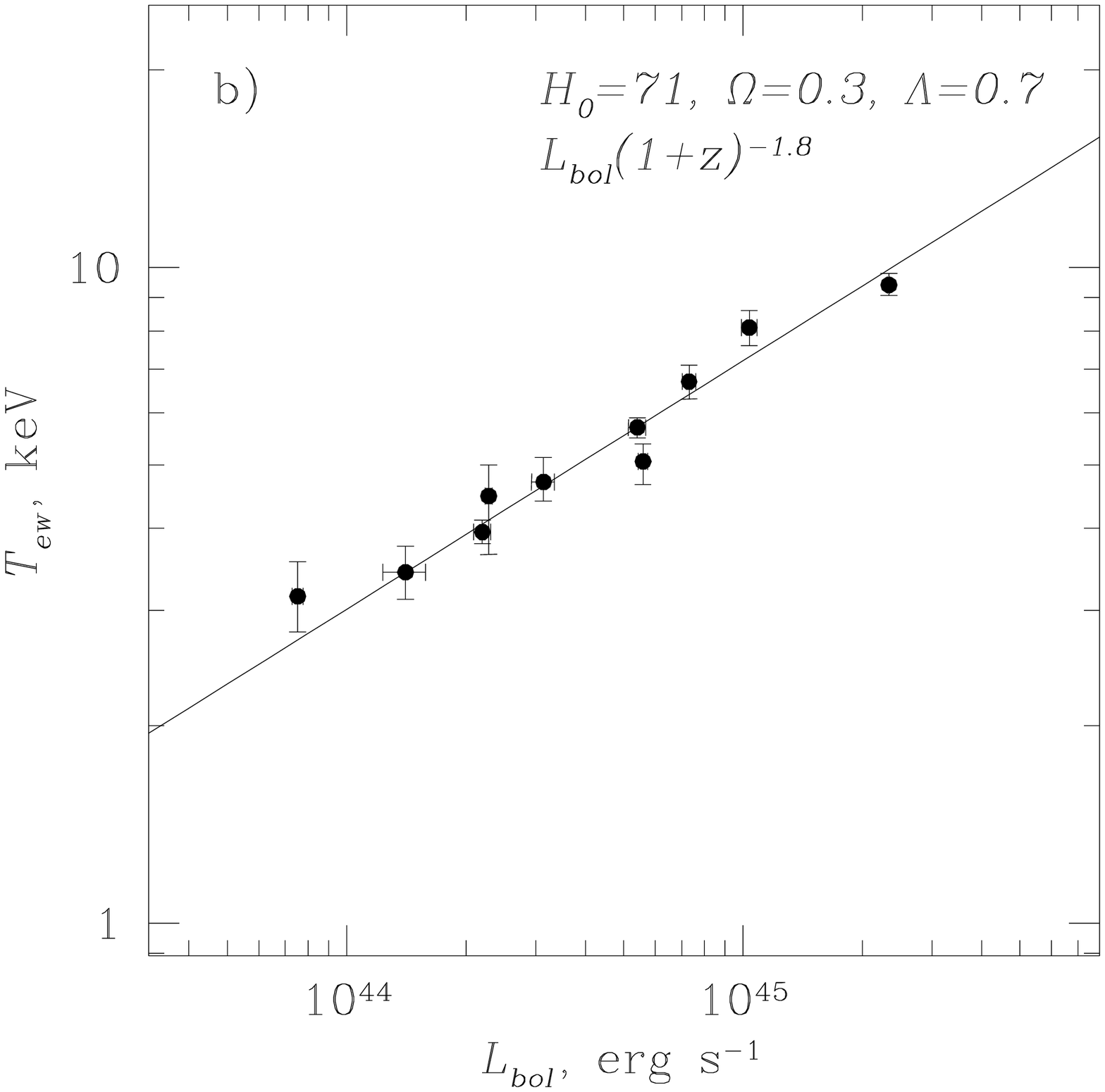}%
}
\caption{Correlation of cluster cooling flow corrected bolometric
  luminosity and $T_{emw}$.  The data are corrected for the {\em
    XMM-Chandra(ASCA)} systematic discrepancies in the temperature and
  flux measurement. The solid line shows the $L-T$ correlation for the
  low-redshift clusters \citep{Markevitch98}. {\em (b)}  shows the
  correlation corrected by the obtained best-fit evolution
  factor,$(1+z)^{-1.8}$ .  
}
\label{fig:lt2}
\end{figure*}

The reference low-redshift $M-T$ relation was adopted from a
\emph{Chandra} sample of \cite{Vikhlinin05a}, which is also close to the
\emph{XMM} results of \cite{astro-ph/0502210}). The \emph{Chandra} sample
contains 13 nearby clusters with exposures sufficient to measure
temperature profiles to $r\approx r_{500}$. The temperatures of these
clusters, 2--10~keV, match well the temperature range in our distant
sample. 
\cite{Vikhlinin05a} derived the mass-temperature relation, $M_{500} = A_5
(T_{spec}/5\,\text{keV})^{1.61\pm0.11}$, where $A_5=(4.13 \pm 0.23)
10^{14} M_{\odot}$. The derived slope is consistent with the
self-similar expectation, $M \propto T^{3/2}$.

Figure \ref{fig:mt}a shows the quantity $a = M_{500}/T_{spec}^{1.5}/A_5$ as a
function of $z$ for our clusters. This quantity represents the
normalization of the $M-T$ relation as constrained by each cluster. For
non-evolving $M-T$ relation, all values of $a$ would be consistent with 1. 
However, the observed values of $a$ clearly indicate evolution. To
quantify the observed evolution, we fitted the data from
Fig.\ref{fig:mt}a with a power law of $E(z)$, $a=E(z)^{-\alpha}$. The
best-fit index is $\alpha=0.88\pm0.23$, where the error bar includes the
uncertainties in low-$z$ normalization of the $M-T$ relation and
our high-$z$ mass measurements. The derived rate of evolution is
consistent with the theoretically expected one, $\alpha=1$. Therefore
we can assume that the normalization of the $M-T$ relation evolves
exactly as $A\propto E(z)^{-1}$. 


Now we can derive slope and  normalization of the $M-T$ relation 
defined by our distant clusters. 
For this, we corrected the mass measurements for evolution by
multiplying them by $E(z)$. Figure \ref{fig:mt}b shows the corrected
cluster masses as a function of temperature. 
The power-law fit, $E(z) M = A_5 (T/5\,\text{keV})^{\gamma}$, to distant clusters 
only\footnote{We used the bisector method modified to
allow for both the measurement uncertainties and intrinsic scatter
\citep{Akritas}.} gives $E(z) M_{500} =
(3.21\pm0.31)(T_{spec}/5\,\text{keV})^{1.79\pm0.19}\times 10^{14} h^{-1} M_{\odot}$.
The obtained value of the slope,$\gamma=1.79\pm0.19$ is 
statistically consistent with the value, $\gamma=1.5$, predicted
by the self-similar theory.



Finally, we note that the results on $M-T$ relation are insensitive to
the absolute calibration of the \emph{XMM} effective area. The main
effect of small calibration errors is to bias the derived temperatures
by a constant factor. It follows from eq.(\ref{eq:def1_m}) that the
normalization of the $M-T$ relation as given by individual clusters,
$A=M_\Delta/\langle T\rangle^{3/2}$, is
\begin{equation}
  \label{eq:m-t-norm:slopes}
  A =
  \frac{(4/3\pi\,\Delta\,\rho_c)^{-1/2}}{(G\,m_p\,\mu)^{3/2}}\;\left(\frac{T}{\langle T\rangle}\right)^{3/2} \,
  \left(-\frac{d \log \rho}{d \log r} - \frac{d \log T}{ d \log r} \right)^{3/2},
\end{equation}
where the last two terms are evaluated at the overdensity radius,
$r_\Delta$. Small calibration errors lead to small changes in estimated
$r_\Delta$ ($\delta r_\Delta/r_\Delta \approx 0.5\,\delta T/T$) but the
normalization of $M-T$ relation is not affected because both density and
temperature are nearly  power-law functions of $r$ in the interesting
range of radii. 

\section{The X-ray temperature - Luminosity correlation}

Our results can also be used to test the evolution in the cluster X-ray
luminosity versus temperature relation. We performed at
spatially-resolved analysis of the temperature and surface brightness
profiles, including the PSF deconvolution. Therefore, we can directly
exclude the contribution of the central cooling regions to both
temperature and luminosity. This significantly reduces the scatter in
the $L-T$ relation \citep{Markevitch98} and thus makes any evolution
more prominent. 

The X-ray  bolometric luminosities were calculated using the Mewe-Kaastra-Liedahl 
plasma emission model. We used the measured best-fit temperatures for three objects with 
irregular X-ray morphology (MS1054.4-0321, \, WARPJ0152.7-1357,and \, CL0939+472) and the obtained 
emission-weighted  temperatures for all other clusters.
All temperatures were renormalized by  $+8\%$ to account for \emph{Chandra} versus \emph{XMM-Newton}  
cross-calibration\footnote{\cite{Vikhlinin02} demonstrated that there is no
bias between \emph{Chandra} and \emph{ASCA} temperatures. Our
\emph{XMM} temperatures are on average 8\% lower than the
\emph{Chandra} values, therefore we need to apply the temperature
 renormalization for a consistent comparison with the low-$z$
\emph{ASCA} measurements by \cite{Markevitch98}.}.
For the  three objects with  irregular X-ray morphology, we use the observed $0.5 - 2.0$ keV 
counts rates within $0<r<1.4$~Mpc  as the normalizing fluxes. For all other clusters,
the normalizing fluxes were calculated by subtracting from the observed $0.5 - 2.0$ keV 
counts rates within $0<r<1.4$~Mpc the fluxes calculated from  the best-fit 
$\alpha$-$\beta$ models within $r \le 70$ kpc and multiplying the result by a factor of
1.06 to account for the flux within $r \le 70$ kpc in a typical $\beta$-model cluster  (see \citep{Markevitch98}).
The resulting luminosities are listed in Table \ref{main2}.

Figure~\ref{fig:lt2} shows the resulting $L-T$ relation. The solid line shows the relation
for low-redshift clusters \citep{Markevitch98}. Clearly, distant
clusters have higher luminosities for the given temperature. 
Parametrizing the evolution as $L=A(1+z)^{\gamma} T_{emw}^{\alpha}$, we
obtain $\gamma=1.8\pm 0.3$ for $\alpha$ fixed at the value for
low-redshift sample, $\alpha=2.64$ (Fig.~\ref{fig:lt2}b). 
The observed rate of evolution is 
consistent with the results of \cite{Vikhlinin02} obtained from an
independent \emph{Chandra} sample. 

\section{Conclusions}

We presented a spatially-resolved analysis of the ICM density and
temperature radial profiles in a sample of 10 $z>0.4$ clusters
observed  with \emph{XMM-Newton}. The main results can be summarized as follows. 

We use the spatially-resolved measurements to study evolution of the
$L-T$ relation excluding the cluster cooling cores. The observed
evolution of the bolometric luminosity for a fixed temperature is
$L\propto(1+z)^{1.8\pm0.3}$ in the redshift range $z=0.4-0.7$, fully
consistent with earlier \emph{Chandra} results \citep{Vikhlinin02} and
other \emph{XMM-Newton} studies \citep{2004A&A...420..853L,maughan05},
but in apparent conflict with conclusions of \cite{2004A&A...417...13E}. 

Most of our clusters have non-constant temperature profiles.
There is a temperature decline at
large radii in most objects, which is qualitatively consistent with the
temperature profiles in low-redshift clusters
\citep{1998ApJ...503...77M,2002ApJ...567..163D,2005ApJ...628..655V}. 

Using the derived temperature and density profiles we determine the
total cluster mass within the radius $r_{500}$ without using the usual
assumption of an isothermal $\beta$-model. This allows a direct
comparison of the high-redshift $M-T$ relation with the recent
high-quality \emph{Chandra} and \emph{XMM-Newton} measurements for
low-redshift clusters (\cite{astro-ph/0502210}; \cite{Vikhlinin05a}). 

The observed $M-T$ relation for distant clusters is $E(z) M_{500} =
(3.21\pm0.31)(T_{spec}/5\,\text{keV})^{1.79\pm0.19}\times 10^{14} h^{-1} M_{\odot}$.
The derived slope, $\gamma=1.79\pm0.19$, is  statistically consistent with
both $\gamma=1.61\pm0.11$ measured by \cite{Vikhlinin05a} for low-$z$
clusters and $\gamma=1.5$ predicted by the self-similar theory.

\acknowledgements

We would like to thank M.~Markevitch and B.~Maughan for useful
discussions.
This work was supported by NASA grant NAG5-9217 and contract NAS8-39073.  O. K. thanks SAO for
hospitality during the course of this research. 

\begin{figure*}
\centerline{\includegraphics[width=0.43\linewidth]{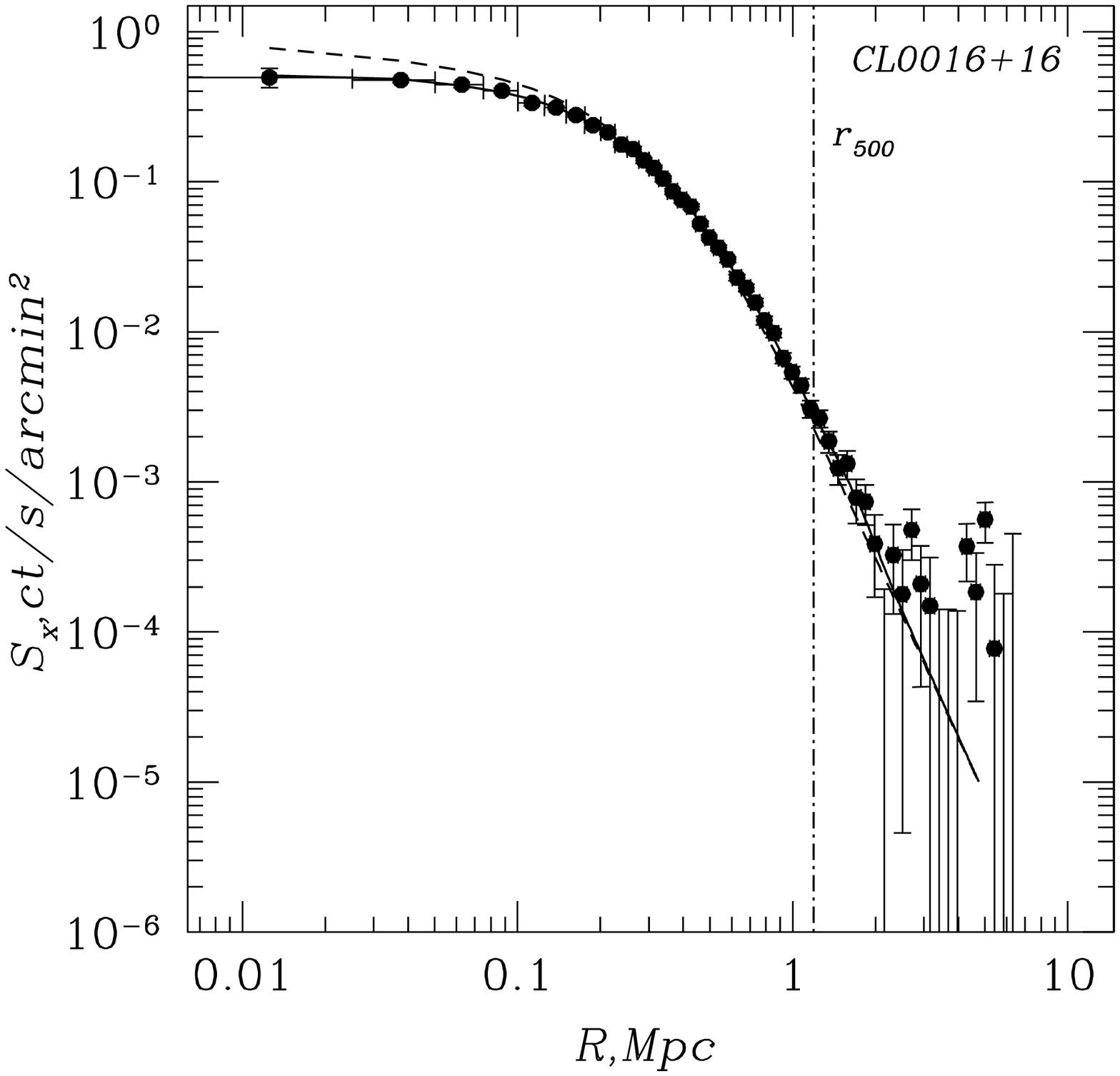}%
\hfill%
\includegraphics[width=0.43\linewidth]{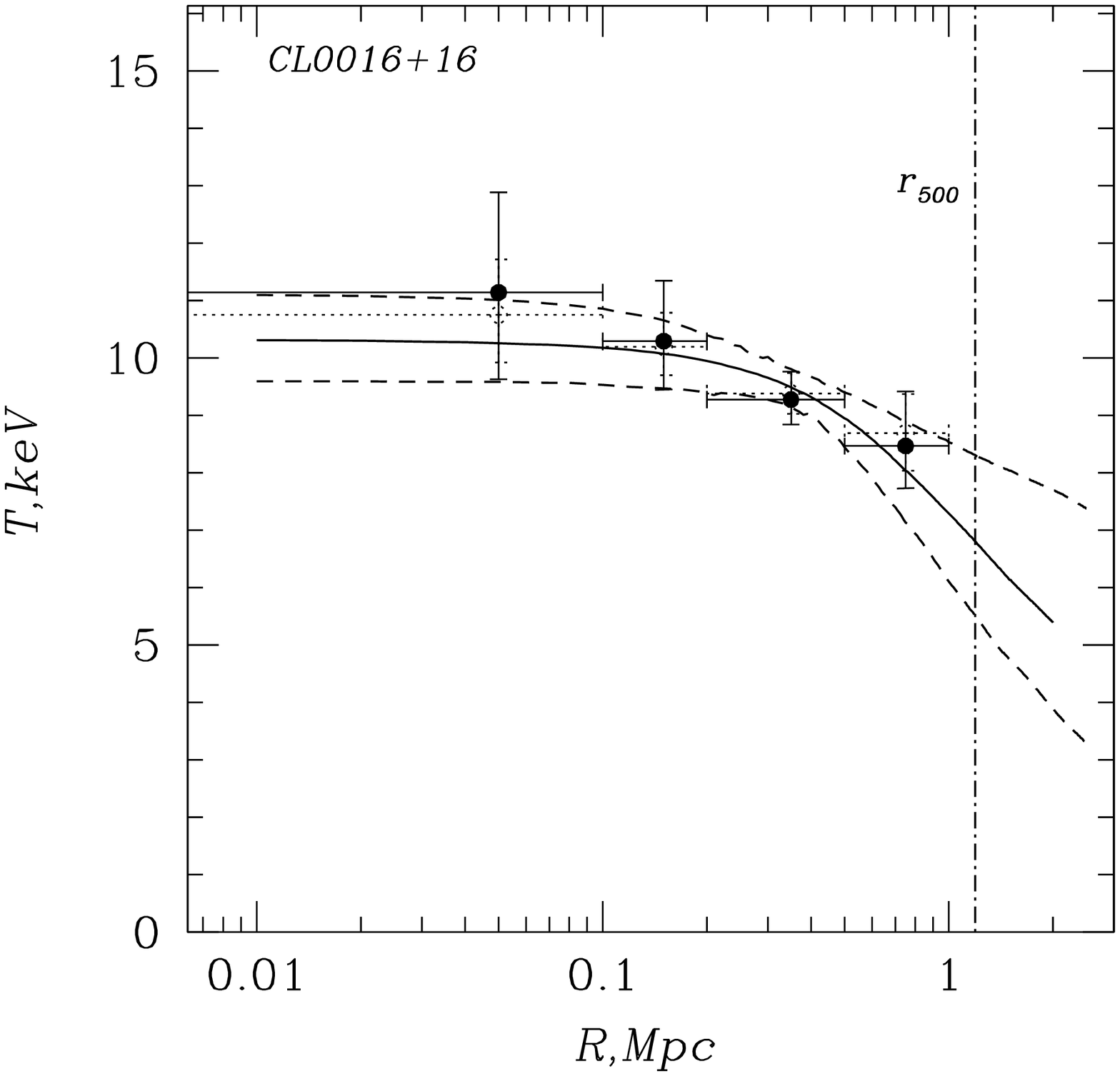}%
}
\caption{\emph{Left:} Observed X-ray surface brightness profile (PN,
MOS1,2 combined) of CL0016+16. Solid line shows the best fit
$\alpha$-$\beta$ model convolved with the PSF. For comparison, dashed
line shows the same model without the PSF degradation. \emph{Right:}
Solid circles show the deconvolved projected temperature profile. For
comparison, open circles show the raw measurements from the X-ray fit
in the same annuli. Solid line shows the best-fit projected
temperature profile and dashed lines correspond to its 68\% CL
uncertainties.} 
\label{fig:cl0016}
\end{figure*}

\begin{figure*}
\centerline{\includegraphics[width=0.43\linewidth]{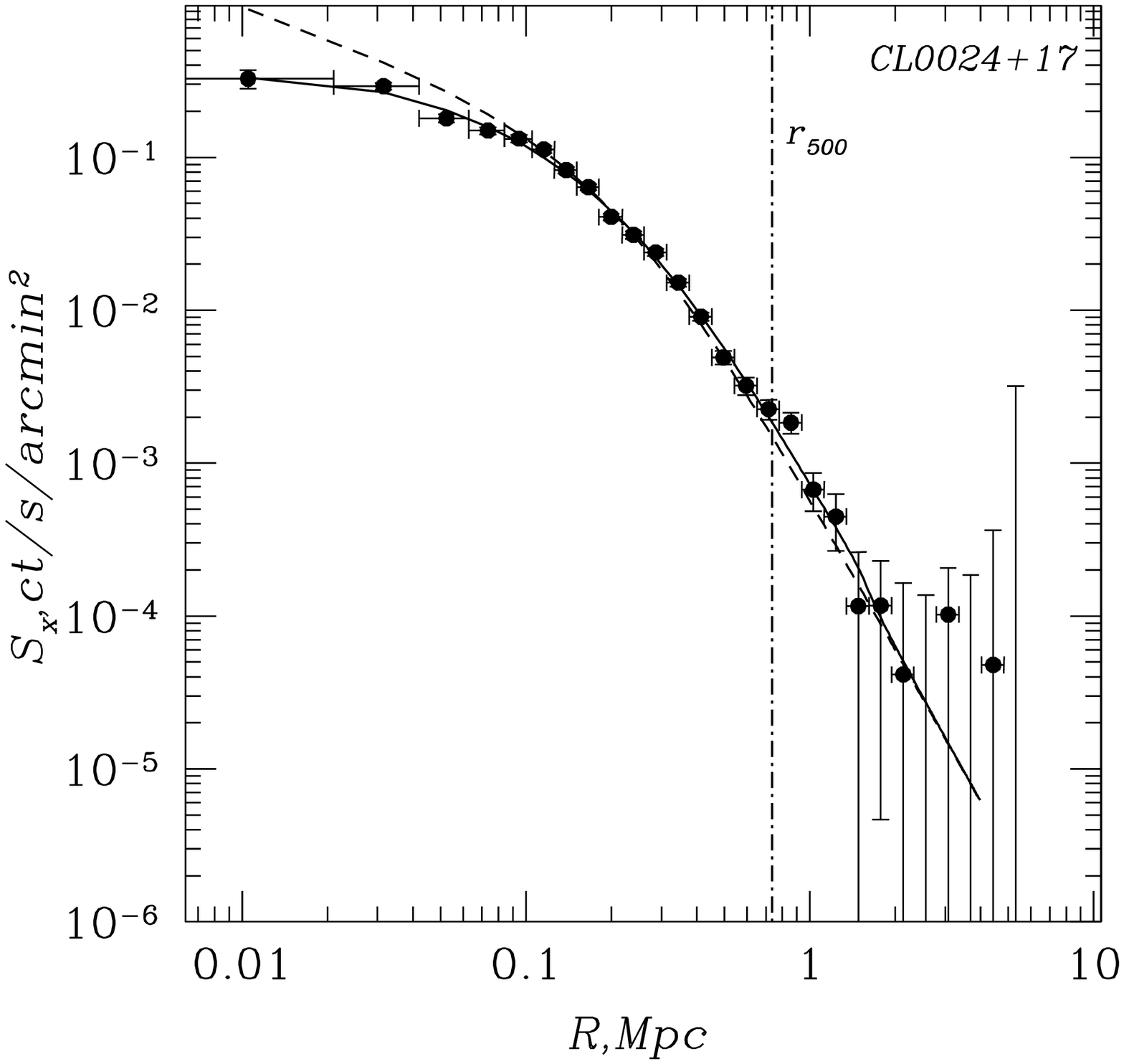}%
  \hfill%
  \includegraphics[width=0.43\linewidth]{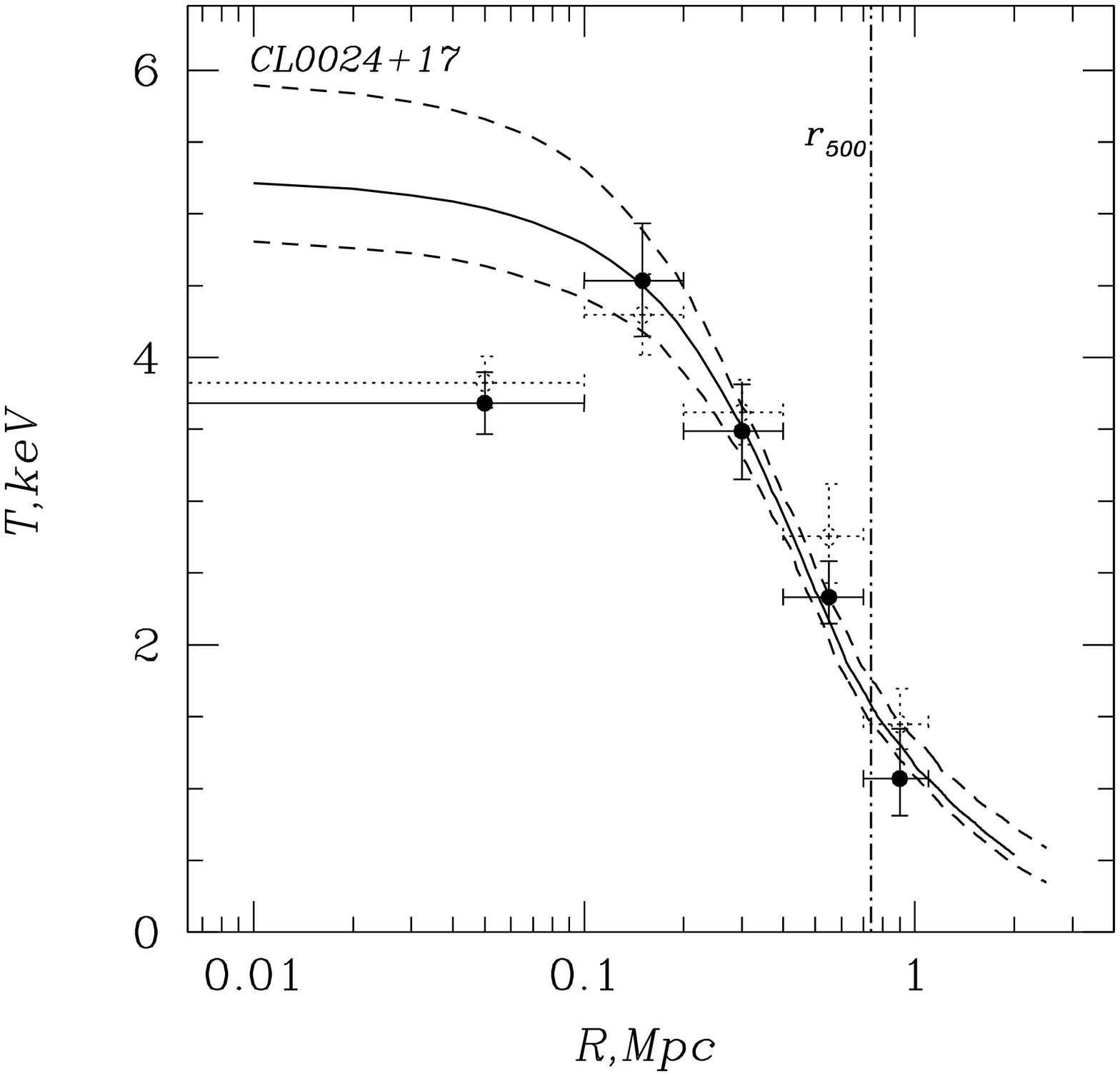}%
}
\caption{Same as Fig.~\ref{fig:cl0016}, but for CL0024+17. 
}
\label{fig:cl0024}
\end{figure*}

\begin{figure*}
\centerline{\includegraphics[width=0.43\linewidth]{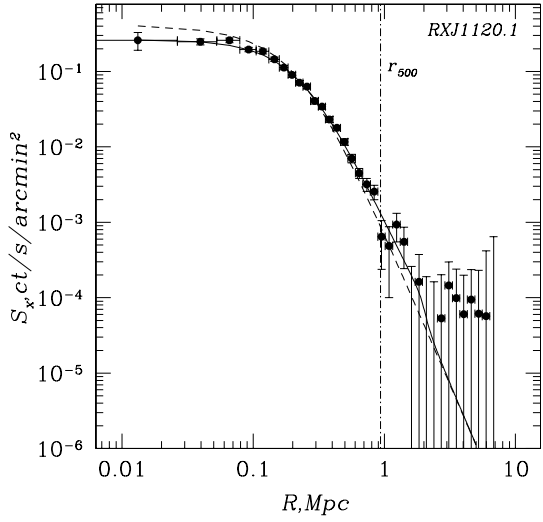}%
  \hfill%
  \includegraphics[width=0.43\linewidth]{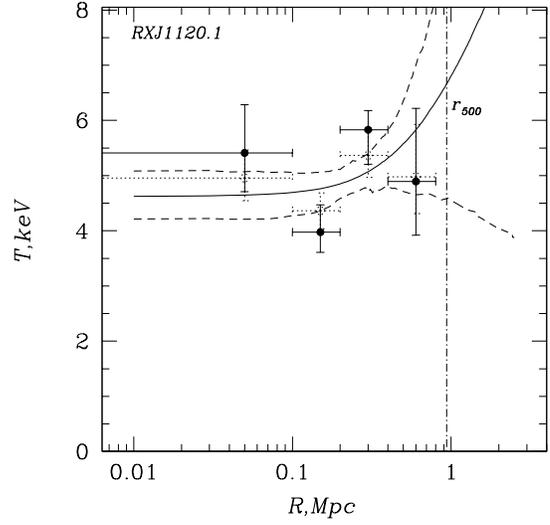}%
}
\caption{Same as Fig.~\ref{fig:cl0016}, but for RXJ1120.1+4318
}
\label{fig:rxj1120}
\end{figure*}

\begin{figure*}
\centerline{\includegraphics[width=0.43\linewidth]{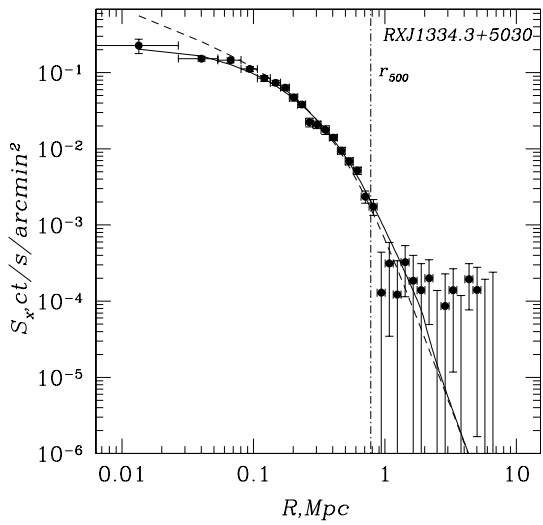}%
  \hfill%
  \includegraphics[width=0.43\linewidth]{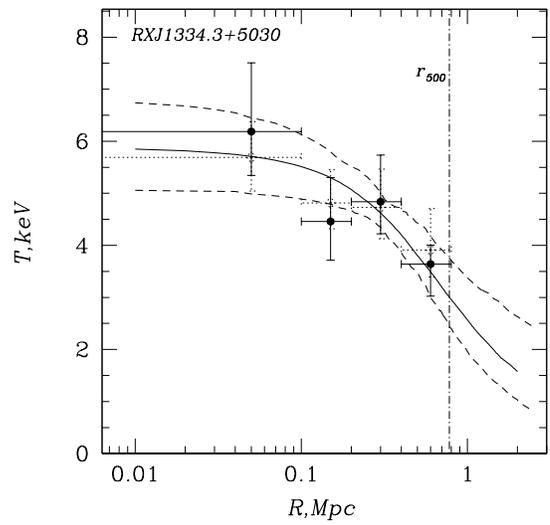}%
}
\caption{Same as Fig.~\ref{fig:cl0016}, but for RXJ1334.3+5030. 
}
\label{fig:rxj1334}
\end{figure*}

\begin{figure*}
\centerline{\includegraphics[width=0.43\linewidth]{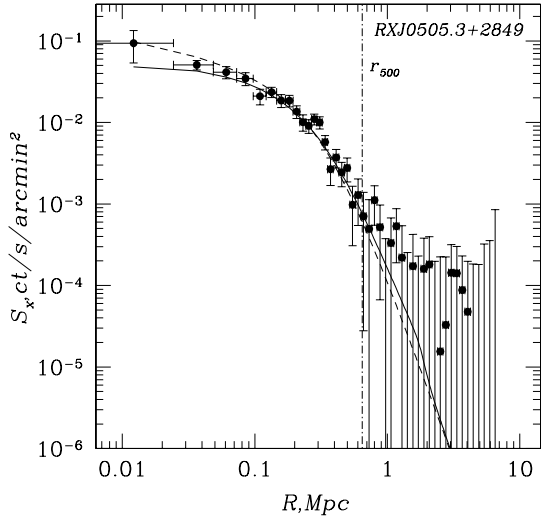}%
  \hfill%
  \includegraphics[width=0.43\linewidth]{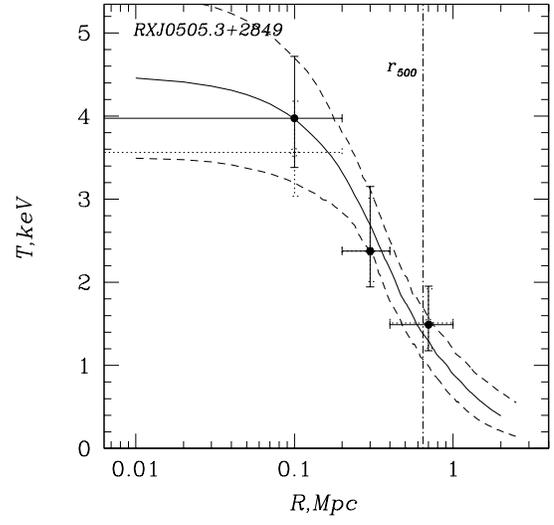}%
}
\caption{Same as Fig.~\ref{fig:cl0016}, but for RXJ0505.3+2849. 
}
\label{fig:rxj0505}
\end{figure*}

\begin{figure*}
\centerline{\includegraphics[width=0.43\linewidth]{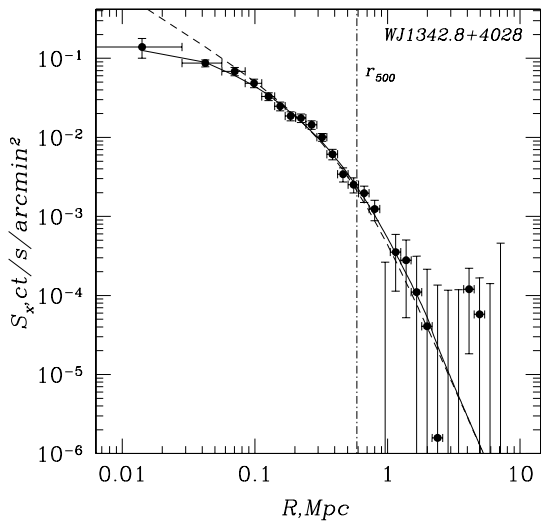}%
  \hfill%
  \includegraphics[width=0.43\linewidth]{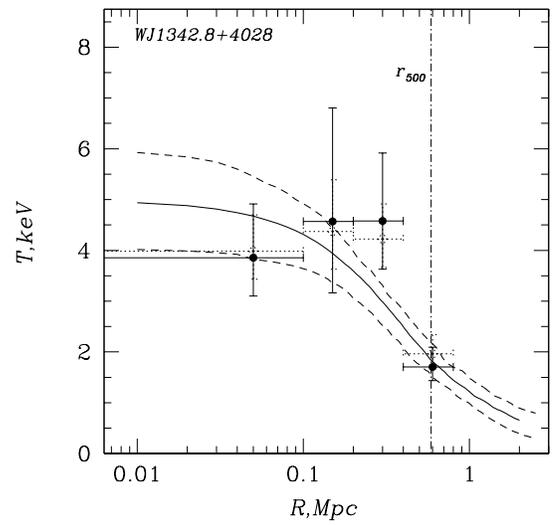}%
}
\caption{Same as Fig.~\ref{fig:cl0016}, but for WJ1342.8+4028. 
}
\label{fig:wj1342}
\end{figure*}

\begin{figure*}
\centerline{\includegraphics[width=0.43\linewidth]{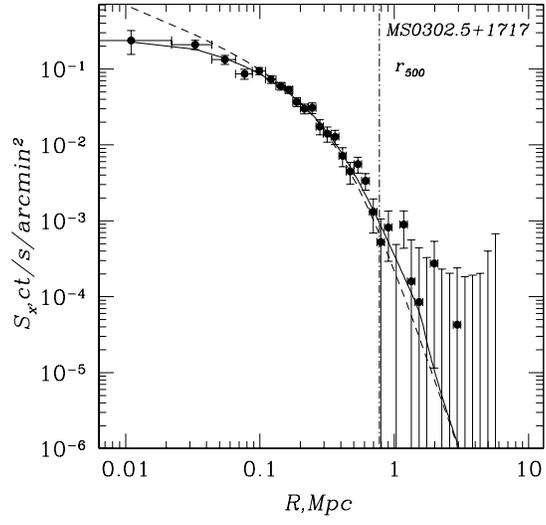}%
  \hfill%
  \includegraphics[width=0.43\linewidth]{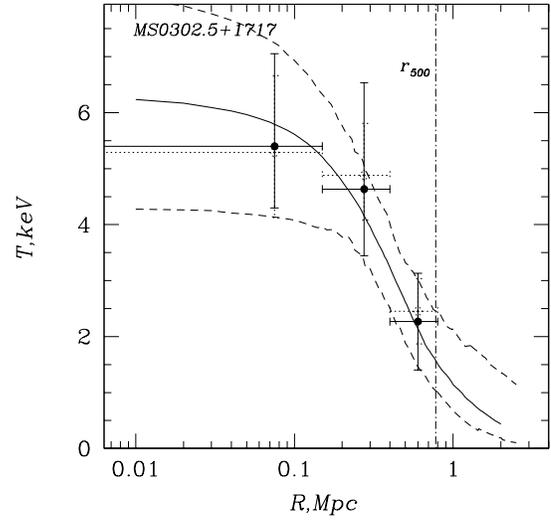}%
}
\caption{Same as Fig.~\ref{fig:cl0016}, but for MS0302.5+1717. 
}
\label{fig:ms0302}
\end{figure*}

\end{document}